\documentclass[prb,twocolumn,floatfix,preprintnumbers,amsmath,amssymb,superscriptaddress]{revtex4}

\usepackage{graphicx}
\usepackage{dcolumn}
\usepackage{amsfonts}
\usepackage{bm}

\begin{document}


\title{Tuning the magnetic ground state of a triangular lattice system}

\author{V. Ovidiu Garlea}
 \email{garleao@ornl.gov}
\affiliation{Neutron Scattering Science Division, Oak Ridge National Laboratory, Oak Ridge, TN 37831, USA}
\author{Andrei T. Savici}
\affiliation{Neutron Scattering Science Division, Oak Ridge National Laboratory, Oak Ridge, TN 37831, USA}
\affiliation{Department of Physics and Astronomy, Johns Hopkins University, Baltimore, MD 21218, USA}
\author{Rongying Jin}
\affiliation{Department of Physics and Astronomy, Louisiana State University, Baton Rouge, LA 70803, USA}


\begin{abstract}
The anisotropic triangular lattice of the crednerite system Cu(Mn$_{1-x}$Cu$_{x}$)O$_{2}$ is used as a basic model for studying the influence of spin disorder on the ground state properties of a two-dimensional frustrated antiferromagnet. Neutron diffraction measurements show that the undoped phase (x=0) undergoes a transition to antiferromagnetic long-range order that is stabilized by a frustration-relieving structural distortion. Small deviation from the stoichiometric composition alters the magnetoelastic characteristics and reduces the effective dimensionality of the magnetic lattice. Upon increasing the doping level, the interlayer coupling changes from antiferromagnetic to ferromagnetic. As the structural distortion is suppressed, the long-range magnetic order is gradually transformed into a two-dimensional order.
\end{abstract}


\maketitle

Magnetic materials with frustrated spin interactions are known to exhibit remarkable sensitivity to small perturbations that could favor certain states among their highly degenerate ground state manifold. Such perturbations can be triggered by thermal fluctuations,~\cite{Villain,Bergman,Bernier} single-ion anisotropy,~\cite{Harris} Dzyaloshinskii-Moriya and dipolar interactions,~\cite{Canals, Palmer} disorder or magnetoelastic coupling,~\cite{Villain2,Ueda,Lee,Ratcliff,Castella,Yamashita,Tchernyshyov,Becca,Saunders,Andreanov} and result in formation of N\'{e}el states, spin glasses or spin liquid states. Among these, perhaps the most intriguing phenomenon is the frustration-relieving lattice distortion and antiferromagnetic ordering resulting from a coupling between the spin and lattice degrees of freedom. This type of lattice instability was found to generically occur in pyrochlore structures,~\cite{Ueda,Lee}) as well as in two-dimensional (2D) triangular systems.~\cite{Melzi,Ye,Giot,Damay,Vecchini} Less emphasis has been placed on the interplay between magnetoelastic coupling and the spin disorder that can produce new exotic phases. Recent theoretical work suggests that a weak exchange randomness tends to stabilize the high-symmetry lattice structure while favoring a spin-glass formation.~\cite{Saunders} While this hypothesis is confirmed experimentally for the pyrochlore lattice (e.g. Zn$_{1-x}$Cd$_{x}$Cr$_{2}$O$_{4}$~\cite{Ratcliff}), it deserves further consideration for the 2D case. At the same time, it is also worth pondering whether there are any intermediate states derived from a diluted magnetic lattice that can be accessed experimentally.

\begin{figure}[btp]
\includegraphics[width=3.5in]{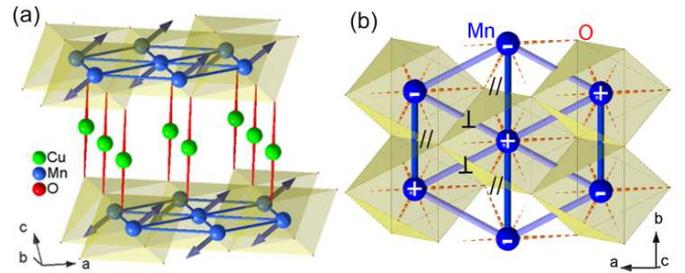}
\caption{\label{structure}(Color online) (a) Stereoscopic view of crystal and magnetic structure of CuMnO$_{2}$ consisting of [Mn$^{3+}$O$_{2}^{2-}$]$^{-}_{\infty}$ layers connected by non-magnetic Cu-planes. (b)Topology of the Mn magnetic lattice in the MnO$_{2}$ layer, defined by two different exchange interactions $J_{\parallel}$ and $J_{\perp}$. }
\end{figure}

We explore such a scenario on the hole-doped crednerite system Cu(Mn$_{1-x}$Cu$_{x}$)O$_{2}$, made of stacked triangular layers of magnetic Mn$^{3+}$ ions positioned in an edge-sharing octahedral coordination. The successive [Mn$^{3+}$O$_{2}^{2-}$]$^{-}_{\infty}$ layers are connected by non-magnetic Cu-planes, where the Cu$^{+}$ is in a linear stick coordination with the oxygens above and below (see Fig.~\ref{structure}(a)). The undoped CuMnO$_{2}$ is unique among its quasi two-dimensional congeners (A$^{+}$B$^{3+}$O$_{2}$ with delafossite structure) in that it displays a monoclinic distortion which arises from a Jahn-Teller type lattice instability.~\cite{Topfer,Trari,Damay,Vecchini} This gives rise, in the $ab$ plane, to a two-dimensional array of Mn edge-sharing isosceles, as illustrated in Fig.~\ref{structure}(b). The in-plane magnetic exchange scheme can be represented by two distinct exchange integrals: $J_{\parallel}$ that defines an AFM chain parallel to the $b$ axis, and $J_\bot$, denoting the frustrated couplings along the two equivalent isosceles sides. The undoped CuMnO$_{2}$ was found to exhibit a magnetoelastic stabilization of a long-range magnetic ordering below approximately 65~K,~\cite{Damay,Vecchini} whereas an earlier study suggested that a small amount of Cu atoms substituting for Mn can cause a suppression of the magnetic order.~\cite{Trari} Below we report neutron diffraction results, which provide valuable insight into the intermediate crystallographic and magnetic states of Cu(Mn$_{1-x}$Cu$_{x}$)O$_{2}$ as a function of composition and temperature. We found that a small deviation from the stoichiometric composition, results to a significant impact on the magnetoelastic properties of the system, and induces a gradual relaxation of the effective dimensionality of the magnetic lattice.

The Cu(Mn$_{1-x}$Cu$_{x}$)O$_{2}$ samples were prepared by solid-state reaction in evacuated sealed quartz tubes at 1000$^\circ$C. Various compositions (x $\simeq$ 0, 0.03, 0.07) were obtained from corresponding stoichiometric mixtures of Cu$_2$O and Mn$_2$O$_3$ oxides. Samples were systematically characterized by x-ray diffraction and magnetization measurements. Neutron powder diffraction experiments were conducted using the HB2A diffractometer at the High Flux Isotope Reactor.~\cite{HB2A} Data was collected using the 1.538~\AA wavelength, produced by a Ge[115] monochromator. Measurements were made on powder samples of approximately 7~g, held in vanadium containers. Rietveld refinements were performed using the Fullprof program.\cite{Fullprof}

The neutron diffraction data confirmed the purity of samples, and also allowed an accurate determination of the Cu/Mn stoichiometry, using the large difference between the their scattering lengths (b$_{coh}$(Mn)~=~-3.75 x 10$^{-13}$cm, and b$_{coh}$(Cu)~=~7.72 x 10$^{-13}$cm). For the room temperature data, the Rietveld refinements were based on the crednerite-type structure model with the monoclinic space group $C2/m$.\cite{Topfer,Trari,Damay,Vecchini} In this structure, the Mn$^{3+}$ and Cu$^+$ occupy the 2$a$ (0 0 0) and 2$d$ (0 $\frac{1}{2}$ $\frac{1}{2}$) Wyckoff positions, while the oxygens are filling the 4$i$ ($x$ 0 $z$) position. The results of the refinements show that the additional Cu ions preferentially substitute for Mn atoms at octahedral site. The refined Mn site occupancies for the studied samples are: 0.992(4), 0.972(4), and 0.932(5), respectively. On the other hand, the oxygen stoichiometry has been determined to be 1.99(1), very close to the nominal value 2. While the Cu ion accommodated in the octahedral site is expected to be bivalent, the electrical neutrality requires the creation of Mn$^{4+}$ ions. The substitution of Mn by Cu leads to a decrease of the $a$ lattice constant and an increase of $\beta$ angle, while the $b$ and $c$ parameters remain almost unchanged. This results from a reduction of the MnO$_6$ octahedron distortion with the Cu substituting for Mn (the apical bond contracts with increasing the doping), which in turn causes a shortening of the ``inter-chain'' Mn-Mn($\bot$) distance.

\begin{figure}[tp]
\includegraphics[width=3.4in]{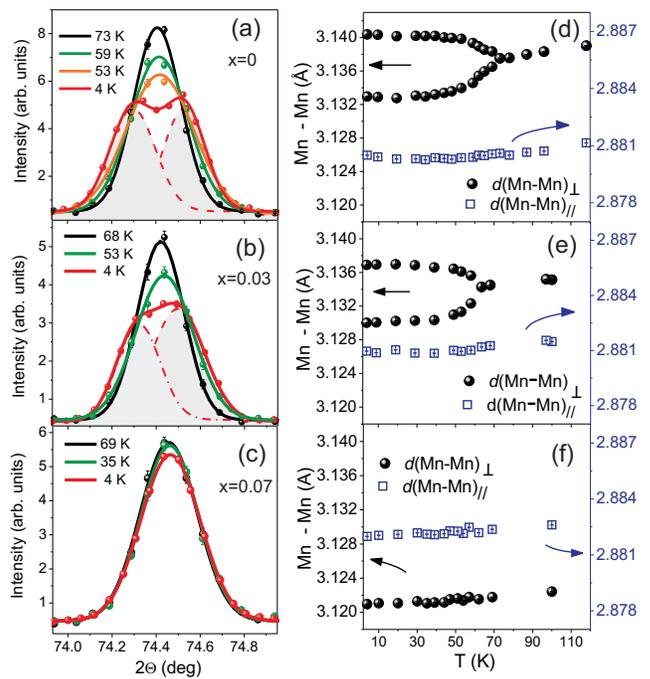}
\caption{\label{splitting}(Color online) (a)(b)(c) Splitting of the (220) monoclinic Bragg peak into two components described by a triclinic unit cell. (d)(e)(f) Temperature dependence of the Mn-Mn bond length for three different compositions. While the substitution of Mn by Cu leads to a shortening of the Mn-Mn($\perp$) distance, there is also a clear trend towards reducing the structural distortion. }
\end{figure}

Similar to previous observation by Damay et \textit{al.},\cite{Damay} the low temperature crystal structure of the parent compound, CuMnO$_{2}$, was found to be triclinic with a slight departure of the $\alpha$ and $\gamma$ from 90$^{\circ}$ ($\alpha$= 90.164(1)$^{\circ}$ and $\gamma$ = 89.833(1)$^{\circ}$, at T = 4~K). This small distortion of the oxygen framework can be accommodated in a pseudo-monoclinic $C\overline{1}$ symmetry, which maintains a similar description of the high-temperature crystal lattice. The occurrence of structural transformation is clearly demonstrated in the Fig.~\ref{splitting}(a) by the splitting of the (220) Bragg peak into two components. The MnO$_6$ octahedra undergoes a small distortion in its equatorial plane, which produces a segregation of Mn-Mn($\bot$) bond into two different components: $d_{\bot1}$ = 3.140(1)\AA~along [110] and $d_{\bot2}$ = 3.132(1)\AA~along [1$\overline{1}$0]. Figure~\ref{splitting}(d) illustrates the evolution of the Mn-Mn bonds length with the temperature, and shows the structural transformation at about 65~K.
Further analysis reveals that for the case of the 3\% Cu-doped sample, the transition takes place at a slightly lower temperature, $\approx$~63~K, and produces a smaller relative change in the bonds length (Fig.~\ref{splitting}(b)(e)). A much pronounced effect is observed for the case of 7\% Cu-doping, where the splitting of the Bragg reflections can no longer be detected (Fig.~\ref{splitting}(c)). The structural refinements for the latter sample were performed assuming a monoclinic cell, identical to that at the room temperature. The results summarized in Fig.~\ref{splitting} clearly show the trend towards reducing the distortion of the frustrated bonds as the Cu-doping increases.

\begin{figure}[tp]
\includegraphics[width=3.2in]{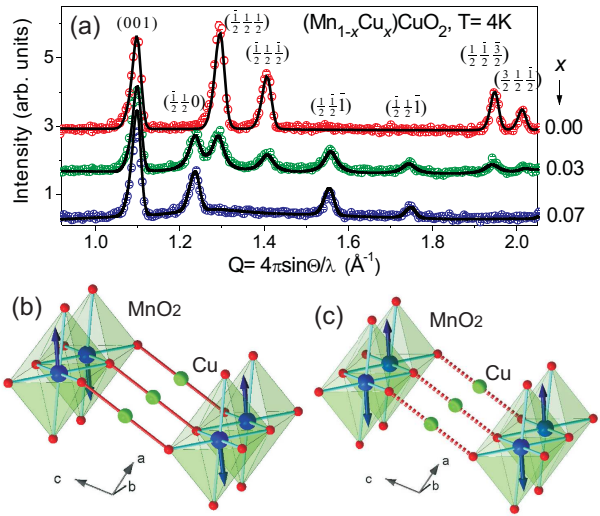}
\caption{\label{diffraction}(Color online) (a) Neutron diffraction data collected for three different Cu(Mn$_{1-x}$Cu$_{x}$)O$_{2}$ powder samples at T = 4~K. The magnetic scattering is indexed by two propagation vectors: ($\frac{\overline{1}}{2},\frac{1}{2},\frac{1}{2}$) and ($\frac{\overline{1}}{2},\frac{1}{2},0$). Solid lines represent the fits to the data. (b),(c) Graphical illustration of the magnetic coupling between successive Mn-layers, changing from AFM in the undoped (b) to FM in the Cu-doped sample (c).}
\end{figure}

The low-temperature structural transition in CuMnO$_{2}$ is found to be closely correlated to the onset of a long range antiferromagnetic order. The magnetic reflections that appear below 65~K were indexed by a propagation vector $k$=($\frac{\overline{1}}{2},\frac{1}{2},\frac{1}{2}$). Interestingly, neutron measurements show that the 3\% Cu-doped sample exhibits two distinct long-range-ordered magnetic phases. A first set of magnetic peaks indexed by $k_1$=($\frac{\overline{1}}{2},\frac{1}{2},\frac{1}{2}$) appears at 63~K, concomitantly with the structural change. This is closely followed, at 58~K, by the formation of another set of magnetic peaks described by $k_2$=($\frac{\overline{1}}{2},\frac{1}{2},0$). The two magnetic phases coexist down to 4~K without changing their relative ratio. Yet, in the case of the 7\% Cu-doped sample, one observes only one set of magnetic Bragg reflections that correspond to the wave vector ($\frac{\overline{1}}{2},\frac{1}{2},0$). This order sets in at approximately 52~K and is not associated with a distortion of the lattice. Figure.~\ref{diffraction}(a) displays a comparison of the base temperature neutron scattering data from all three samples.

Spin configurations compatible with the crystal symmetry were generated by group theory analysis using the program SARA$h$.\cite{Sarah} The model that best fits the data consists of Mn moments aligned in the $(ac)$ plane, pointing along the direction of the ferrro-ordered $d_{3r^2-z^2}$ orbital. The spin structure is depicted in Fig.~\ref{structure}. One of the most interesting and somewhat surprising findings of this work is the change in the magnetic coupling along the $c$ direction between the stacked antiferromagnetic layers. While in the parent compound the adjacent MnO$_2$ layers are coupled antiferromagnetically, in the doped samples we observe a gradual conversion to a ferromagnetic alignment as the level of Cu-doping increases. In the 7\% Cu doped sample the coupling between layers becomes purely ferromagnetic. This seems to be a direct consequence of the spin-disorder, i.e., it cannot be explained by the structure change as the linear interplanar O-Cu-O bonds are slightly shrinking with increasing $x$. A stereoscopic view of the magnetic coupling between octahedral layers is displayed in Fig.~\ref{diffraction}(b)(c). Our experimental results clearly indicate a tendency towards vanishing the 3D long-range ordered moment, as the Mn magnetic lattice gets randomly diluted with Cu ions. Rietveld fits of the 4~K data yielded an ordered moment $m_{Mn}\approx3.03(5)~\mu _{B}$ for the CuMnO$_{2}$ and 1.89(8)~$\mu _{B}$ for Cu(Mn$_{0.93}$Cu$_{0.07}$)O$_{2}$. For the intermediate doping (3\% Cu), both magnetic phases described by the wave vectors ($\frac{\overline{1}}{2},\frac{1}{2},\frac{1}{2}$) and ($\frac{\overline{1}}{2},\frac{1}{2},0$) exhibit ordered moments of similar magnitudes 2.46(10)$~\mu _{B}$ and 2.49(8)$~\mu _{B}$, respectively.

\begin{figure}[tp]
\includegraphics[width=3.4in]{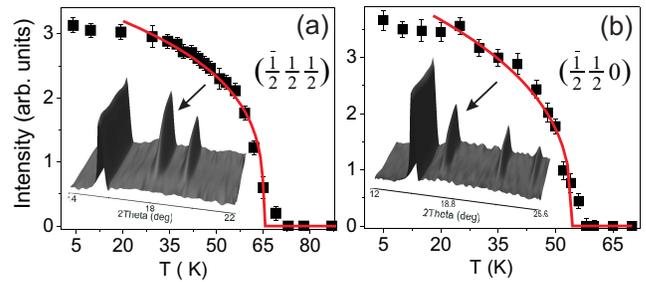}
\caption{\label{ordparam}(Color online) Variation with the temperature of the measured neutron scattering for (a) undoped and (b) 7\%-Cu doped samples. The temperature dependence of magnetic peak intensities was fit using a power-law model.}
\end{figure}

Figure~\ref{ordparam} shows the temperature dependence of the magnetic peaks integrated intensities observed for CuMnO$_{2}$ and Cu(Mn$_{0.93}$Cu$_{0.07}$)O$_{2}$. Besides the shift to a lower temperature of the transition point, the intensities follow similar trends. To assess the character of the long-range order transition, we measured the critical exponent $\beta$ associated with the magnetic order parameter defined as $m(T)\propto\sqrt{I}\propto(T_{N}-T)^{\beta}$. Least-square fits to the data near the transition ($\Delta T\approx~$30~K) yielded an exponent $\beta\simeq$~0.14(1) for the undoped, and $\simeq$~0.13(3) for the 7\%-doped sample. Although the number of data points is somewhat limited, and likely affected by the presence of critical scattering, the estimated values of the critical exponent $\beta$ are close to those of the 2D Ising ($\beta=0.125$) universality class. This is certainly not surprising considering the 2D character of the crystal structure.

\begin{figure}[]
\includegraphics[width=3.3in]{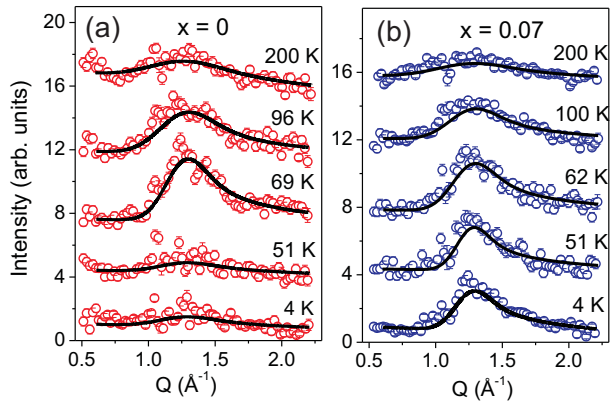}
\caption{\label{diffuse}(Color online) Temperature dependence of the magnetic diffuse scattering in Cu(Mn$_{1-x}$Cu$_{x}$)O$_{2}$ with x=0 (a) and x=0.07 (b). Solid lines represent the fit of the scattering signal using a 2D Warren function, as described in the text.}
\end{figure}

In addition to the magnetic Bragg peaks, one also observes broad asymmetric scattering around the ($\frac{\overline{1}}{2},\frac{1}{2},0$) peak position, that persists in a very wide range of temperatures, starting from above 200~K and down to 4~K. This is due to the critical scattering associated with a 2D short range spin correlation, and appears to become progressively stronger when approaching the N\'{e}el temperature. Figure~\ref{diffuse} shows the evolution with the temperature of the 2D scattering profile, obtained by subtracting the calculated diffraction pattern assuming a flat background. The powder averaged magnetic intensity of ($hk$) reflections of the 2D lattice is described as:
\begin{equation}
I_{hk}\propto \frac{|F_{m}(\mathbf{Q})|^{2}}{\sin\theta}\int_0^{\frac{\pi}{2}}\exp\left({-\frac{4\pi\xi^2}{\lambda^2}(\sin\theta \cos\varphi-\sin\theta_{hk})^2}\right)\mathrm{d}\varphi
\nonumber
\end{equation}
where $F_{m}(\mathbf{Q})$ is the magnetic structure factor, $\xi$ is the correlation length of the magnetic ions, $\lambda$ is the neutrons wavelength and $\theta_{hk}$ represents the angular position of magnetic reflections ($hk$).\cite{Warren,Wu} The estimated correlation length for all measured samples is $\xi\approx$ 5\AA~at 200~K, and 14 \AA~in the proximity of N\'{e}el temperature. However, unlike the undoped sample where the 2D correlations are gradually giving way to the 3D long range order, the doped samples show no decrease in the 2D phase fraction. Instead, in the absence of structural distortion, the short-range magnetic order competes with and prevails over the 3D order. This behavior is qualitatively similar to the spin freezing in the pyrochlore lattice, suggested to be generated by random strains in the presence of weak magnetoelastic coupling.\cite{Saunders,Ratcliff}

In summary, the neutron diffraction study of triangular lattice systems Cu(Mn$_{1-x}$Cu$_{x}$)O$_{2}$, reveals a very rich spectrum of ground states as a function of temperature and chemical doping. Our measurements confirmed that the stoichiometric phase CuMnO$_{2}$ undergoes a magnetoelastic transition from high-temperature monoclinic-paramagnetic to low-temperature triclinic-antiferromagnetic phase. Upon substituting Cu for Mn, the structural distortion is gradually relaxed along with the antiferromagnetic coupling between adjacent MnO$_2$-layers. For a 7\%-Cu doping, the exchange randomness tends to stabilize a 3D magnetic phase with ferromagnetic interplanar coupling defined by the wave vector ($\frac{\overline{1}}{2},\frac{1}{2},0$). As the structural distortion decreases, short-range 2D correlations that develop above T$_N$, start to build up over the long range order and persist down to the low temperature region.

After the completion of the manuscript, we became aware of the paper by Poienar \emph{et al.},\cite{Poienar} which presents a study of Cu$_{1.04}$Mn$_{0.96}$O$_{2}$. The compound is reported to exhibit a frustration-lifting distortion and 3D order with ferromagnetic interplanar coupling.

The authors thank B. Chakoumakos, C. Broholm and D. Singh, for valuable discussions and interest in this work.
V.O.G. is grateful to B. Sales for making available his laboratory for samples preparation.
Work at ORNL was supported by the U.S. Department of Energy (DOE) under Contract No. DE-AC05-00OR22725 with UT-Battelle, LLC. R.J. is supported by NSF DMR-1002622.

\end{document}